\documentclass{sig-alternate-10pt}
\usepackage[utf8]{inputenc}
\usepackage[T1]{fontenc}
\usepackage{graphicx,url}
\usepackage{colortbl}
\usepackage{caption}
\usepackage{subcaption}
\usepackage{subfig} 
\usepackage{float}
\usepackage{wrapfig}
\usepackage{multirow}
\usepackage{balance}

\newcommand{\footnoteref}[1]{\textsuperscript{\ref{#1}}}

\begin{document}
%\title{Factors that Influence the Formation of Discriminatory Female Stereotypes in Search Engines}
%\title{The contribution of global images to the formation of physical attractiveness stereotypes in search engines}
%\title{The contribution of global images to the formation of stereotypes in search engines}
%\title{Exploring factors that lead to  bias\\in search engine image results for physical attractiveness}

\title{\vspace{-10pt}Stereotypes in Search Engine Results: Understanding The Role of Local and Global Factors\vspace{-10pt}}
%\title{Analyzing Local and Global Influence in the Creation of Stereotypes in Search Engine Answers}
\author{ 
\numberofauthors{3} 
\alignauthor
Gabriel Magno\\
       \affaddr{Universidade Federal de Minas Gerais}\\
       \affaddr{Belo Horizonte, Brazil}\\
       \email{magno@dcc.ufmg.br}
% 2nd. author
\alignauthor
Camila Souza Ara\'{u}jo\\
       \affaddr{Universidade Federal de Minas Gerais}\\
       \affaddr{Belo Horizonte, Brazil}\\
       \email{camilaaraujo@dcc.ufmg.br}
% 3rd. author
\alignauthor 
Wagner Meira Jr.\\
       \affaddr{Universidade Federal de Minas Gerais}\\
       \affaddr{Belo Horizonte, Brazil}\\
       \email{meira@dcc.ufmg.br}
\and  % use '\and' if you need 'another row' of author names
% 4th. author
\alignauthor Virgilio Almeida\titlenote{Computer Science Department at UFMG(virgilio@dcc.ufmg.br)}\\
       \affaddr{Berkman Klein Center, Harvard University}\\
       \affaddr{Cambridge, USA}\\
       \email{valmeida@cyber.law.harvard.edu}
}
\maketitle

\begin{abstract}
The internet has been blurring the lines between local and global cultures, affecting in different ways the perception of people about themselves and others. In the global context of the internet, search engine platforms are a key mediator between individuals and information. In this paper, we examine the local and global impact of the internet on the formation of female physical attractiveness stereotypes in search engine results. By investigating datasets of images collected from two major search engines in 42 countries, we identify a significant fraction of replicated  images. We find that common images are clustered around countries with the same language. We also show that existence of common images among countries is practically eliminated when the queries are limited to  local sites. In summary, we show evidence that results from search engines are biased towards the language used to query the system, which leads to certain attractiveness stereotypes that are often quite different from the majority of the female population of the country.
\end{abstract}

\vspace{-3pt}
\section{Introduction and Motivation}

All over the world, search engines are powerful mediators between individuals and the access to information and knowledge. General search engines play a major role when it comes to give visibility to cultural, social and economic aspects of the daily life~\cite{Anthes2015}.  Recent studies have demonstrated that the ranking of answers provided by search engines have a strong impact on individuals  attitudes, preference and behavior~\cite{Epstein2015}. Usually, people  trust the answers in higher ranks, without having any idea how the answers get ranked by  complex and opaque algorithms~\cite{Pasquale2015}. 

Search engines can be viewed as part of a broad class of social algorithms, that are used  to size us up, evaluate what we want,  and provide a customized experience~\cite{Lazer2015}. Physical attractiveness  is a pervasive and powerful agent in the social world, that is also being affected  by social algorithms and by the growing digitization of the physical world.  Physical attractiveness has influence on decisions, opportunities and perceptions of ourselves and others. So, one natural question arises: what is the impact of search engines on the perception of physical attractiveness?  Our previous work on search results  identified stereotypes for female attractiveness in images available in the Web\cite{araujo2016identifying}. 

Stereotypes can be regarded as ``pictures in our head that portray all members of a group as having the same attribute''~\cite{greenwald2016}. They are generally defined as beliefs about the characteristics, attributes, and behaviors of members of certain groups~\cite{Hilton1996}. As pointed out in~\cite{greenwald2016}, humans think with the aid of categories and categories are the basis for normal prejudgment. In many circumstances, categories turn into stereotypes, such as Africans have rhythm or Asians are good at math.  Stereotypes may also be associated with some prejudgment, that  indicates some sort of social bias, positive or negative. Age, race, gender, ethnicity, and  sexual orientation are elements that contribute to the creation of stereotypes in different cultures.

Stereotypes can evolve in ways that are linked to social and cultural changes. Some stereotypes and prejudgment found in the material world are transferred to the online world. For example, ~\cite{kay2015unequal}  identified gender stereotypes in image search results for occupation. Considering the internet is blurring the lines between local and global cultures, a relevant question is to understand the impact of local and global factors on the formation of stereotypes in the internet. The mechanism of repetition (e.g., repetition of music, videos, images, etc.) is one step that characterizes the influence of  globalization on local cultures. This work aims at  understanding  the role of local and global factors on the formation of stereotypes found in search engine results for physical attractiveness.

In order to understand the local and global impact of internet on stereotypes, we focus on the analysis of answers provided by search engines in different countries to questions associated with physical attractiveness. 
The complexity of internet search platforms, such as Google and Bing, makes it impossible to look for transparency of their algorithms and data. So, our  approach for the stereotype problem is to follow the concept of transparency of inputs and outputs (aka as black-box techniques) of a class of specific 
 queries~\cite{Chander2016}. This type of approach has been successfully used to analyze the behavior of complex systems, such as virtual machines~\cite{Wood2007}. 
Black-box techniques infer information about the behavior of systems by simply observing each virtual machine from the outside and without any knowledge of the application resident within each machine. 
Several interesting observations related to bias and fairness were learned from the quantitative analysis of the global and local answers provided by the search engines to our set of input queries on female physical attractiveness.

\vspace{-3pt}
\section{Background and Related Work}

%In this section we present some related work on characterization studies of search engines, stereotypes, as well as physical attractiveness.

This section discusses some references that deal with search engine characterization, stereotypes and discrimination.
In some specific situations,  search engines may show biased answers. Therefore it is important to be able to  understand how the result ranking is built and how it affects the access to information \cite{introna2000shaping}. \cite{noble2013google} shows how racial and gender identities may be misrepresented, when commercial  interest is involved. \cite{umoja2012missed} has questioned  commercial search engines because the way they represent women, especially black women and other marginalized groups. This type of behavior masks and perpetuates unequal access to social, political and economic life of some groups. 
 
Stereotyping can be viewed as oversimplified ideas about social groups, it reduces a person or thing to those traits while exaggerating them \cite{baker2013white}. Stereotypes can be positive, neutral or negative. A recent study by Kay et al. \cite{kay2015unequal} shows a systematic under representation of women in image search results for occupations. This kind of stereotype affects people’s ideas about professional gender ratios in the real world and may create conditions for bias and discrimination. In \cite{Sweeney2013} the author shows that Google searches involving  names suggestive of race are more likely to serve up arrest-related ads indicating signs of discrimination.

\begin{figure*}[!ht]
\centering
\vspace{-15pt}
\includegraphics[width=0.8\textwidth]{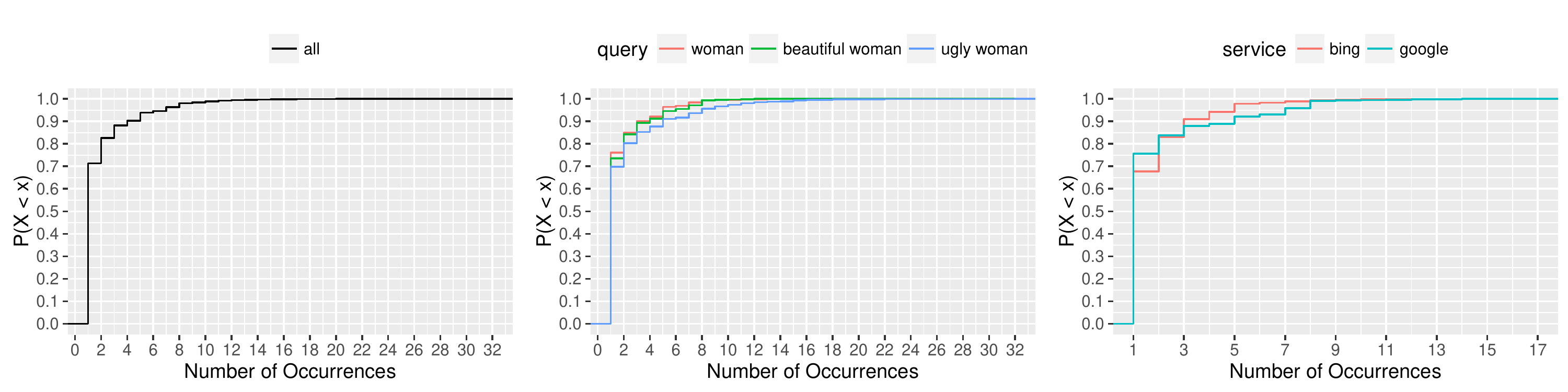}
\vspace{-6pt}
\caption{CDF of image repetition.}
\label{fig:imgs_repetition_cdf}
\vspace{-6pt}
\end{figure*}

\vspace{-3pt}
\section{Methodology}

This section presents the methodology for gathering and analyzing data. Our methodology aims to identify the influence of globalization of the internet and local culture  on the formation of stereotypes through two factors: language and location. The starting point of our analysis is a set of image queries, defined by the context of interest (in our case attractiveness stereotypes) submitted to different search engines. We then analyze, for each query, the top 100 images checking which images do repeat across queries as well as image characteristics (e.g., race) and try to draw patterns that arise for languages and countries. In particular, 
since the same language may be spoken in several countries, we employ a two-level strategy, where we first check for patterns at the language granularity and then we also consider location as well. 

In the following sections, we first describe the data gathering strategy, then the procedure to generate image fingerprints that will allow to detect the occurrence of the same image in several queries and finally the similarity metric used to compare query results.

\subsection{Data Gathering}

Data gathering was carried through two search engine APIs for images: Google\footnote{https://developers.google.com/custom-search/} and Bing\footnote{https://www.microsoft.com/cognitive-services/en-us/bing-image-search-api}. Once gathered, we extract features from the images using Face++\footnote{http://www.faceplusplus.com/}. In summary, the data gathering process consists of:

\begin{small}
\begin{enumerate}
\item {\bf Define search queries}\\
Translate each search query - beautiful woman, ugly woman and woman - to the target languages \footnote{Using Google Translator}.
\item {\bf Gathering}\\Using the search engine APIs, perform the searches for the defined queries in the  countries of our lists. Afterwards, we remove any images that contain no faces or multiple faces.
\item {\bf Extract attributes}\\Identify faces and infer race using the face detection tool.
\end{enumerate}
\end{small}

We then build two different datasets, one with default parameters and the other with parameters to return only results of the same country. For both datasets, each query is associated with a single country, that is, it is expressed in the official language of the country and submitted to a service whose address is in the top level domain (TLD) of the target country. The first dataset, named global, does not restrict the source of the images in terms of TLD of the site that provides them, that is, the images collected are not necessarily from hosts in the country for which the API is submitting the search. The second dataset is named local, since we also define the country from which the images must come.

Using the APIs we were able to obtain 100 images for query, but we consider as valid only images in which Face++ was able to detect a single face. The analysis will be performed for all query responses that contain at least 20 valid images. The three query searches (beautiful woman, ugly woman and woman) were performed for several countries, providing a good coverage in terms of regions and internet usage, and their official languages:

\begin{small}
\begin{description}
\item [BING] (total of 5.824 valid images): Saudi Arabia, Denmark, Austria, Germany, Greece, Australia, Canada, United Kingdom, USA, South Africa, Argentina, Spain, Mexico, Finland, Italy, Japan, South Korea, Brazil, Portugal, Russia, Turkey and Ukraine.

\item [GOOGLE] (total of 11.314 valid images): Algeria, Saudi Arabia, Egypt, Iraq, Morocco, Denmark, South Africa, Australia, Canada, United Kingdom, Nigeria, USA, Zambia, Finland, France, Austria, Germany, Greece, India, Ireland, Italy, Japan, South Korea, Malaysia, Afghanistan, Brazil, Portugal, Angola, Russia, Argentina, Chile, Guatemala, Mexico, Paraguay, Peru, Spain, Venezuela, Kenya, Sweden, Turkey, Ukraine and Uzbekistan.
\end{description}
\end{small}

\subsection{Image Fingerprinting}

In order to identify the co-occurrence of images across datasets, we need a method that is able to identify whether two images are the same or not. Matching their URLs is not good enough, since the same image may be provided by different sites. Also, using a hash function such as MD5 or SHA-1 does not solve the problem either, since a re-sized image would be associated with a completely different hash value compared to the original one. 

Ideally, the technique should be able to ``fingerprint'' an image, i.e., to determine a label that uniquely identifies the image, despite small modifications. We use the dHash (difference hash) algorithm~\cite{dHash2013}, which consists of four main steps: (1) it shrinks the image to 9x8 pixels; (2) it converts the image to grayscale; (3) it computes the difference between adjacent pixels; and (4) it assigns bits whenever the left pixel is brighter than the right pixel. This algorithm will output a 64-bit hash value per image that we use to uniquely identify the images in our datasets.

\subsection{Similarity Metric}

An adequate comparison of sets of images returned by a query requires a similarity metric. Given two lists of images, $A$ and $B$, the Jaccard index measures the similarity (or diversity) between $A$ and $B$, and is calculated as $J(A,B) = \frac{|A \cap B|}{|A \cup B|}.$ In other words, it is the ratio between the size of the intersection and the size of the union of $A$ and $B$. The closer the index is to 0, the more diverse the sets are, while an index closer to 1 indicates that $A$ and $B$ are similar.
In practice, each set of images returned by a search is represented as a set of fingerprints, and we determine the similarity of two searches through their Jaccard index.

\vspace{-3pt}
\section{Experiments and Results}

\begin{figure*}
    \centering
    \vspace{-15pt}
    \begin{subfigure}[b]{0.45\textwidth}
        \includegraphics[width=0.8\textwidth]{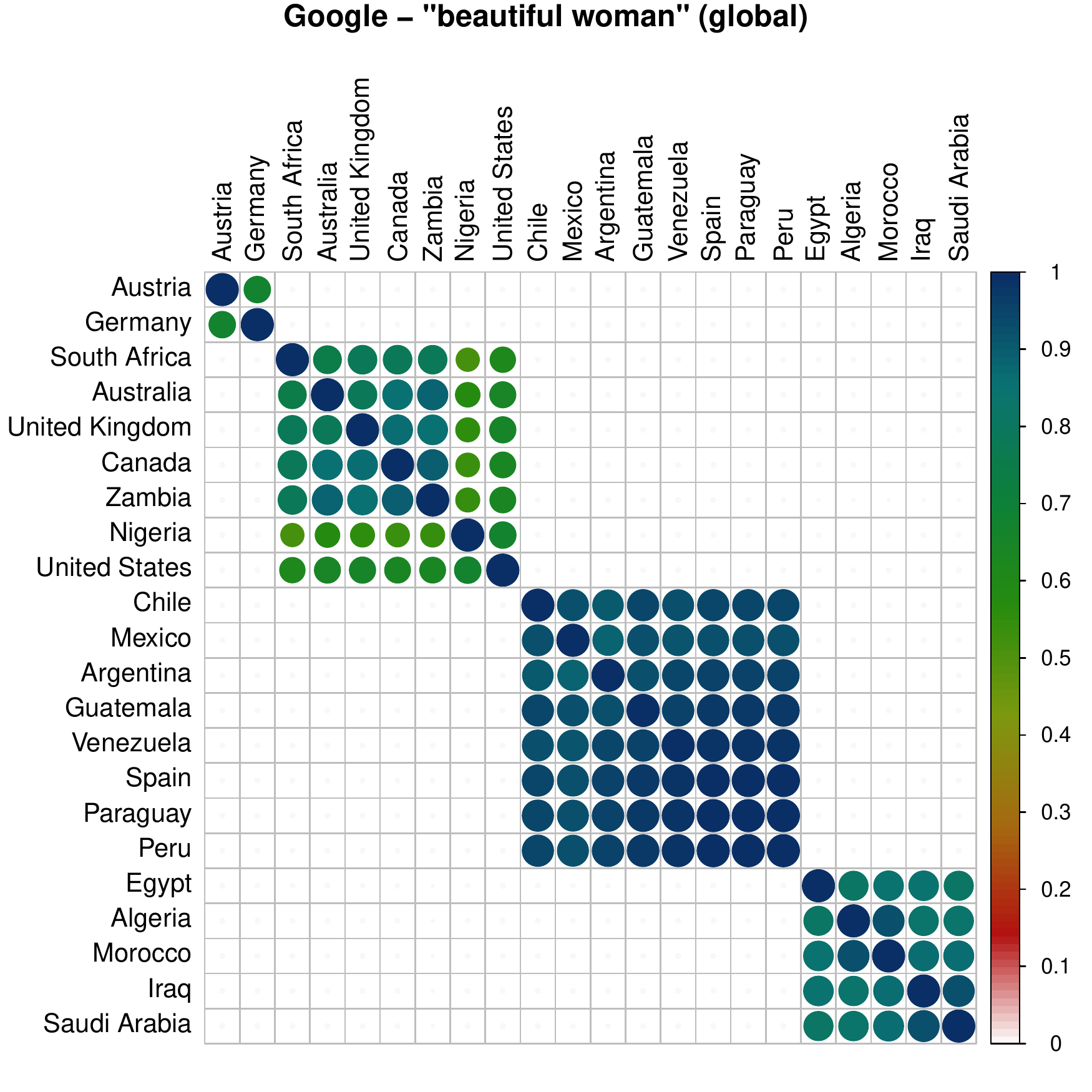}
        %\caption{\hfill}
        %\label{fig:similarity_global}
    \end{subfigure}
    \begin{subfigure}[b]{0.45\textwidth}
        \includegraphics[width=0.8\textwidth]{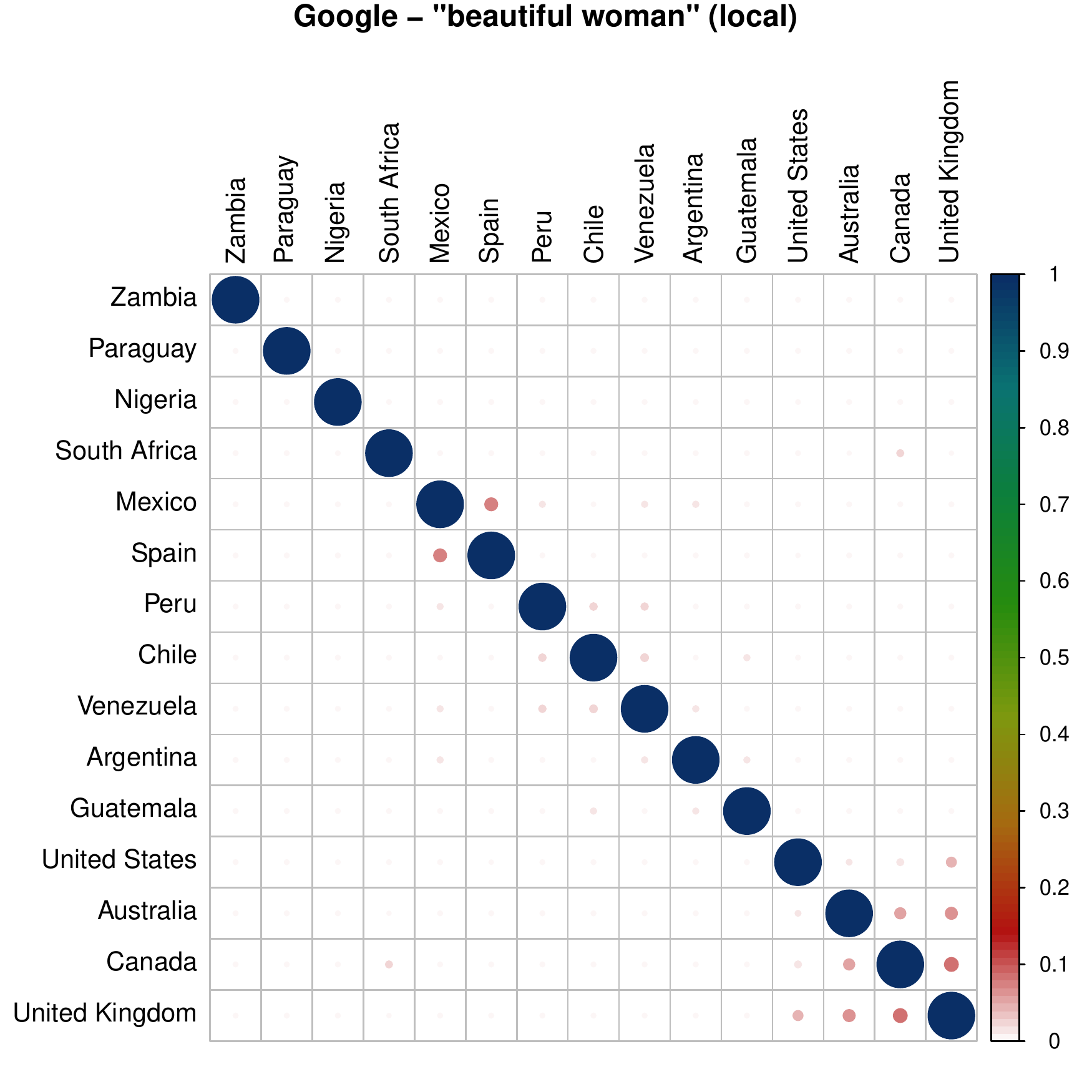}
        %\caption{\hfill}
        %\label{fig:similarity_local}
    \end{subfigure}
    \vspace{-6pt}
    \caption{Similarity of image results between countries, for the query ``beautiful woman'' in Google}\label{fig:similarity_matrix}
    \label{fig:similarityGlobalLocal}
    \vspace{-7pt}
\end{figure*}

This section describes the experiments carried out in our analysis and present the main results. First, we present evidence that images co-occur in different datasets. Then, we characterize the repetition of images across search results by analyzing the similarities between them. Finally, we compare global and local results, analyzing them in terms of similarity and racial profile of the target countries.

\subsection{Repetition of Images}

In order to analyze the repetition of images across our search results, we start by calculating the dHash of each image and determine the frequency of each unique hash value in our datasets. Our goal is to analyze how frequently the same images appear in multiple queries, countries and services. For this experiment we use only the global dataset. 

%\begin{figure*}[!ht]
%\includegraphics[width=\textwidth]{imgs/imgs-repetition-%segmented-count.pdf}
%\caption{Frequency of the number of occurrences (repetition) %of images in our datasets.}
%\label{fig:imgs_repetition}
%\end{figure*}

Figure~\ref{fig:imgs_repetition_cdf} shows the Cumulative Distribution Function (CDF) of the number of repeated images, for three scenarios: whole dataset (left), grouping by query (center) and grouping by service (right). First, we observe that there are, indeed, images that do appear in several sets of results. Although ~71\% of the images are unique, some images appear in up to 33 different sets of results.

Another interesting finding is that images resulting for the query ``ugly woman'' seem to repeat more often than the other queries. For instance, the maximum value of repetition for ``ugly woman'' is 32, whereas for ``beautiful woman'' is 13 and for ``woman'' is 14. 
%Also, analyzing the distribution in Figure~\ref{fig:imgs_repetition_cdf} (center) we observe that ~99\% of the images repeat themselves less than 8 times for plain and beautiful woman, while for ugly the same figure is 96\%.

Comparing the distribution between services, we observe that they are slightly different. In Bing results, ~68\% of the images are unique, while in Google it is ~76\%. 
%Although the percentage of unique images from Google is higher, it also presents a higher fraction of images that repeat compared to Bing: ~7\% of images repeat 6 times or more, compared to only ~2\% for Bing.
These results motivate us to investigate what the factors that influence image repetition are.

\subsection{Similarities}

Now we aim to investigate the reasons for the co-occurrence of images. We measure similarity between services, queries and countries. For the analysis presented in this section just the global dataset is used.

%\vspace{-3pt}
\subsubsection{Services}

In this experiment we analyze the co-occurrence of images in both Bing and Google. We do that by comparing the pairs of image sets (one from Bing and one from Google) for the same query and same country. In this case, there are 22 pairs of services for each query, totaling 66 pairs. 

The average Jaccard indices for plain, beautiful and ugly woman queries are, respectively, $0.04$, $0.06$ and $0.11$, indicating that there is no significant match between results from Bing and Google. Despite that, the similarity for ``ugly woman'' is almost twice as large as the others (on average), supporting our previous finding that ``ugly woman'' images repeat more often.

%We calculate the average and standard deviation Jaccard Index per query, presented in Table~\ref{tab:simil_services}.

%\begin{table}[ht]
%\caption{Similarity between Google and Bing}
%\label{tab:simil_services}
%\vspace{-10pt}
%\centering
%\scriptsize
%\begin{tabular}{l|rr|}
%\cline{2-3}
%              & \multicolumn{2}{c|}{Jaccard Index} \\
%\hline
%\multicolumn{1}{|l|}{Query}           & Avg. & Std. \\ 
%\hline
%\multicolumn{1}{|l|}{woman}           & 0.04 & 0.03 \\ 
%\multicolumn{1}{|l|}{beautiful woman} & 0.06 & 0.04 \\ 
%\multicolumn{1}{|l|}{ugly woman}      & 0.11 & 0.05 \\ 
%\hline
%\end{tabular}
%\end{table}

%\vspace{-3pt}
\subsubsection{Queries}

Analogously to the comparison between services, we will now analyze the co-occurrence of images between queries (e.g "woman" vs. "beautiful woman"). For this scenario we have three possible pairs: "woman" vs. "beautiful woman", "woman" vs. "ugly woman" and "beautiful woman" vs. "ugly woman". Since we compare within each service and country, we analyze 64 pairs ($22 + 42$) for each combination of queries, totaling 192 pairs. 

We observe that, again, the similarity is small. The average Jaccard index for ``ugly woman'' compared to either ``woman'' or ``beautiful woman'' is 0.01 ($std=0.02$). Interestingly, the similarity between ``woman'' and ``beautiful woman'' is three times larger than the other combinations ($avg=0.03$, $std=0.03$), indicating that the plain query (``woman'') tends to give results closer to ``beautiful woman''. It is important to notice that this is a preliminary result, since the standard deviation values are high and the confidence intervals overlap with the average values of the other. 
\vfill

%In Table~\ref{tab:simil_queries}, we present the average and standard deviation values per query configuration.

%\begin{table}[ht]
%\centering
%\scriptsize
%\caption{Similarity between combination of queries}
%\label{tab:simil_queries}
%\vspace{-10pt}
%\begin{tabular}{ll|rr|}
%\cline{3-4}
%        &      & \multicolumn{2}{c|}{Jaccard Index} \\
%\hline
%\multicolumn{1}{|l}{Query 1} & Query 2            & Avg. & Std. \\ 
%  \hline
%\multicolumn{1}{|l}{woman} & beautiful Woman      & 0.03 & 0.03 \\ 
%\multicolumn{1}{|l}{woman} & ugly Woman           & 0.01 & 0.02 \\ 
%\multicolumn{1}{|l}{beautiful woman} & ugly Woman & 0.01 & 0.02 \\ 
%   \hline
%\end{tabular}
%\end{table}

\begin{figure*}[!ht]
\centering
\vspace{-20pt}
\includegraphics[width=0.9\textwidth]{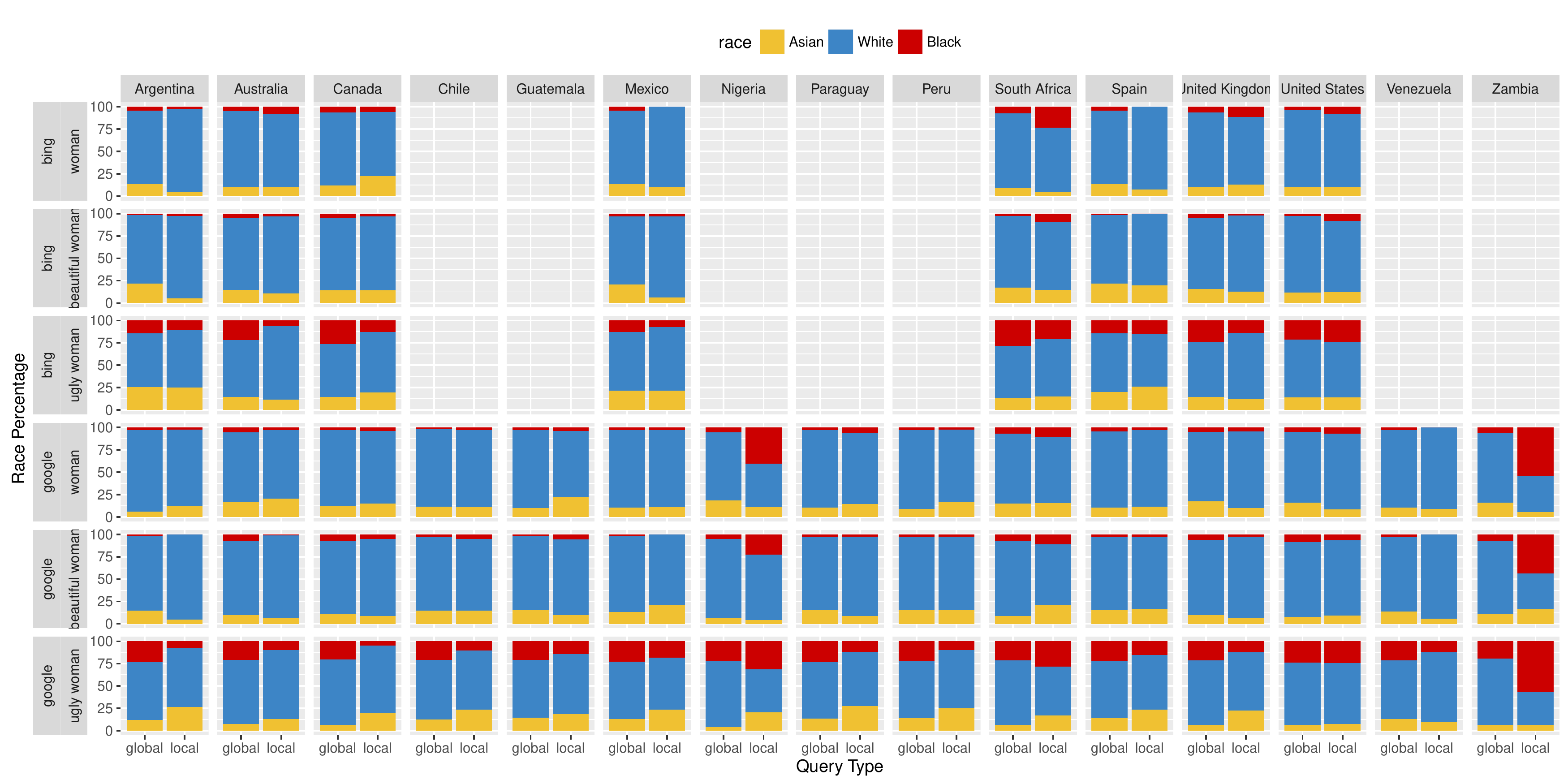}
\vspace{-10pt}
\caption{Distribution of races among countries, queries and services.}
\label{fig:racial_profile}
\vspace{-14pt}
\end{figure*}

\subsubsection{Countries}
\label{sec:similarity_countries}

Finally, we compare the lists between each pair of countries (861 for Google and 231 for Bing), and  calculate their Jaccard index. Due to space limitations, we present here only the results for the query ``beautiful woman'' in Google, but the results are similar for the other datasets and queries. Figure~\ref{fig:similarityGlobalLocal} (left) shows the similarity matrix between countries. To enhance visibility, we present only the 22 countries that cluster with other countries.

In contrast to the service and query analyses, there are very strong similarities between countries. We observe that the similarities are stronger among countries that speak the same language, and almost nonexistent between countries that speak different languages. The influence of language is so pronounced that we may easily identify  ``language-based clusters''.

Such result is explained by the fact that images are indexed by the search engine using the content of the web-page with which the image is associated. Since the queries are issued using written natural language, it is possible that an image returned, for example, by Google Mexico is actually from a site in Spain (e.g., xyz.es)

\subsection{Global and Local Images}

As shown in the previous section, there are very strong similarities between countries. Our hypothesis is that the results of image searches, on both search engine platforms, are biased in relation to language and do not always reflect the characteristics of the female population of the country. 

We investigate the effect of filtering the search query to return only results from a given country, defined by local sites existing in the country code domain of the specific country. For this investigation we select the countries of the two largest clusters (English and Spanish), totaling 8 countries in Bing and 15 in Google. We then collect the images using the same methodology used for searching globally (without the country filter).

%\vspace{-3pt}
\subsubsection{Similarity}

We initially  assess the impact on the similarity between countries when searching images locally. Similarly to Section~\ref{sec:similarity_countries}, we calculate the Jaccard index for each pair of countries. 

Figure~\ref{fig:similarityGlobalLocal} (right) shows the similarity matrix for the local search results. Compared to the matrix for global queries (left), it is visible how the similarity is drastically reduced. The clusters have virtually disappeared, despite some small values ($< 0.1$) remained for the Spanish cluster (Mexico and Spain) and the English cluster (Australia, Canada and United Kingdom).

This result supports our observation that the similarity is almost non-existent between countries that speak different languages. On the other hand, we may easily identify  ``language-based clusters''.

\subsubsection{Racial Profile}

In our previous work \cite{araujo2016identifying}, we have demonstrated the existence of stereotypes for female physical attractiveness, in particular negative stereotypes about black women and positive stereotypes about white women in terms of physical attractiveness. In this work we show how the racial profile of the countries changes when we filter local results, indicating that query results do not reflect the local demography. We then compare the racial distribution of a country when issuing global queries vs. local queries.

It is possible to observe how the racial distribution changes for almost every country/query when the search query is local (Figure~\ref{fig:racial_profile}). For African countries (Nigeria, South Africa and Zambia) the proportion of black women increases for almost all queries - only for 'ugly woman', on Bing, the proportion decreases for local search in South Africa. This result is consistent with the demographics of those countries where most of the population is black~\footnote{\label{note}http://www.indexmundi.com}. On the other hand, the proportion of black women decreases for almost all the local searches in Argentina and Australia, where 97\%~\footnote{https://www.cia.gov/library/publications/the-world-factbook/fields/2075.html} and 92\%
\footnoteref{note}) of the population is white, respectively.

\vspace{-5pt}
\section{Findings and Conclusions}

In this work we study the impact of local and global images on the formation of female physical attractiveness stereotypes. We start by 
analyzing the co-occurren\-ce of images returned by search engines in the context of pictures of women. We queried and downloaded thousands of images from different search engines (Google and Bing), distinct queries (woman, beautiful woman and ugly woman), originally provided to 42 different countries. We showed that repetition occurs across our datasets, and it is more pronounced for ``ugly woman'' pictures. By comparing and calculating the similarity metric between pairs of search results we found out that images between services and between queries tend to differ, while images between countries present very high similarity for countries that speak the same language, forming ``language clusters''. When submitting local queries we observe that the similarity between countries is nearly eliminated. Also, querying locally gives us a more trustworthy racial profile in some cases, reflecting the actual demographics of those particular countries.  Our findings highlight and evidence the fact that results from search engines are biased towards the language used to query the system, which may impose certain stereotypes that are often very different from the majority of the female population of the country. Furthermore, our methodology for investigating search engine bias by analyzing only the input and output is a contribution by itself.

%\section*{Acknowledgments}

%This work was partially funded by Fapemig, CNPq, CAPES, and by projects InWeb, MASWeb, and EUBra-BIGSEA.

\clearpage

\bibliographystyle{abbrv}
\balance
\bibliography{main}

\end{document}